\documentclass{aa}

\usepackage{graphics}

\newcommand{\kmprs}  {\mbox{\rm km\,s$^{-1}$}}

\newcommand{\feh} {\mbox{\rm [Fe/H]}}

\newcommand{\teff}  {\mbox{$T_{\rm eff}$}}
\newcommand{\logteff} {\mbox{${\rm log} T_{\rm eff}$}}
\newcommand{\logg}  {\mbox{{\rm log}$g$}}

\newcommand{\Lione} {\ion{Li}{i}}
\newcommand{\Lisix} {\element[][6]{Li}}
\newcommand{\Liseven} {\element[][7]{Li}}
\newcommand{\sixseven} {\element[][6]{Li}/\element[][7]{Li}}

\newcommand{\Feone} {Fe\,{\sc i}}

\begin{document}

\thesaurus{07(08.01.1, 08.01.3, 08.06.3, 08.09.3, 09.03.2)}

\title{Isotopic lithium abundances in five metal-poor disk stars
\thanks{Based on observations carried out at the European 
Southern Observatory, La Silla, Chile}}

\author{P.E.~Nissen \inst{1} \and D.L.~Lambert \inst{2}
\and F.~Primas \inst{3} \and V.V.~Smith \inst{4}}

\offprints{P.E.~Nissen}

\institute{Institute of Physics and Astronomy, University of Aarhus, DK--8000
Aarhus C, Denmark
\and Department of Astronomy, University of Texas, Austin, TX 78712--1083
\and European Southern Observatory, Karl-Schwarzschild Str. 2,
D--85748 Garching b. M\"{u}nchen
\and Department of Physics, University of Texas at El Paso, El Paso,
TX 79968-0515}

\date{Received 15 April 1999 / Accepted June 1 1999}

\maketitle

\begin{abstract}
High resolution ($R \simeq 110\,000$), very high S/N spectra
centered on the 6707.8~\AA\ \Lione\ line 
have been obtained with the ESO Coud\'{e} Echelle Spectrometer
for five, metal-poor ($-0.8 < \feh < -0.6$) disk stars in the turnoff 
region of the HR-diagram. The instrumental and stellar atmospheric 
line broadening have been determined from two unblended iron
lines and used in a model atmosphere synthesis of the
profile of the \Lione\ line as a function of the lithium isotope 
ratio. This has led to a detection
of \Lisix\ in \object{HD\,68284} and \object{HD\,130551} with
$\sixseven \simeq 0.05$, whereas the other stars,
\object{HR\,2883}, \object{HR\,3578} and \object{HR\,8181},
have \sixseven\ close to zero. 

By comparing \teff -values and absolute magnitudes based on Hipparcos
parallaxes with recent stellar evolutionary tracks, 
the masses of the stars have been derived. It is shown that the
two stars with \Lisix\ present have a significantly higher mass,
${\cal M}/{\cal M}_{\sun} \simeq 1.05$,
than the other three stars for which values between 
0.85 and 1.0 ${\cal M}_{\sun}$ are obtained.

The results are discussed in terms of models for the galactic
evolution of the light elements and depletion of the lithium
isotopes in stellar envelopes. It is shown that the measured \Lisix\
abundances are in agreement with standard cosmic ray production of 
\Lisix\ in the galactic disk and a moderate depletion (0.5 dex) in
the stars. Recent models for the evolution of \Lisix\ including
$\alpha + \alpha$ fusion reactions and predicting a high lithium
isotopic ratio, $\sixseven \simeq 0.3$ at $\feh = -0.6$, require
a high degree of \Lisix\ depletion ($\simeq 1.0$ dex) to fit the observations.
Furthermore, these models imply a \Liseven\ abundance about 0.2 dex 
higher than observed for metal-poor disk stars.

 
\end{abstract}

\section{Introduction}
Measurements of the abundance of the \Lisix\ isotope in stellar
atmospheres
are of considerable interest and have attracted much attention
since the first detection of \Lisix\ in the metal-poor
turnoff star \object{HD\,84937} by Smith et al. (\cite{smith93}).
The reason for this interest is threefold:

{\em i)} Detection of \Lisix\ in halo turnoff stars puts strong limits
on the possible depletion of \Liseven ,
and thus allows better determination of the primordial \Liseven\
abundance from the observed Li abundance of stars on the
`Spite plateau'. 
(Copi et al. \cite{copi97}, Pinsonneault et al. \cite{pin98}) 

{\em ii)}  \Lisix\ abundances as a function of
\feh\ provide an additional test of theories for the
production of the light elements Li, Be and B
by interactions between fast nuclei and ambient ones
(Ramaty et al. \cite{ramaty96}, Yoshii et al. \cite{yoshii97},
Fields \& Olive \cite{fields99}, Vangioni-Flam et al. \cite{flam99}). 

{\em iii)} Information on depletion of \Lisix\  as a function of stellar
mass and metallicity puts new constraints on stellar models
in addition to those set by \Liseven\ depletion. This is so because
the proton capture cross section of \Lisix\ is much larger than that
of \Liseven . Hence, at a given metallicity
there will be a mass interval, where \Lisix\ but not \Liseven\
is being destroyed according to standard stellar models
(Chaboyer \cite{cha94}).

Altogether, \Lisix\ abundances may contribute to the study of
such different fields as
Big Bang nucleosynthesis, cosmic ray physics and stellar structure.
It will, however, require a rather large data set of \Lisix\
abundances to get information in all these areas.
The most metal-poor stars around the turnoff are of particular
interest in connection with the determination of the primordial
\Liseven\ abundance, whereas more metal-rich halo stars and disk
stars are of interest for the study of the formation and astration of
the light elements.

Recent studies of \Lisix\ abundances have concentrated on halo
stars. Following the
first detection of \Lisix\ in \object{HD\,84937} by Smith et al.
(\cite{smith93}) at a level corresponding to an isotopic ratio
of $\sixseven \simeq 0.05$, Hobbs \& Thorburn (\cite{hobbs94}, \cite{hobbs97})
have confirmed the detection, and found upper limits of \sixseven\
for 10 stars. More recently, Smith et al.
(\cite{smith98}) report the probable detection of \Lisix\
in another halo star \object{BD\,+26\,3578}  with about the same
metal abundance, mass and evolutionary stage as \object{HD\,84937},
and give tight upper limits of \sixseven\ for 7 additional stars.
Finally, Cayrel et al. (\cite{cayrel99a}) have observed
the \Lione\ line in \object{HD\,84937} with very high S/N and
confirmed the presence of \Lisix\ with a high degree of confidence.

In the case of disk stars there has not been any
systematic search for \Lisix\ since the studies of
Andersen et al. (\cite{and84}) and Maurice et al. (\cite{mau84}).
In these papers an upper limit of \sixseven\ of about 0.10
is set for about 10 disk stars ranging in metallicity from $-1.0$ to +0.3. 
The meteoritic \sixseven\ ratio is close to 0.08 (Anders \& Grevesse
\cite{anders89}) and
the interstellar ratio is similar -- possibly with significant variations
(Lemoine et al. \cite{lemoine95}). For metal-poor disk stars
the ratio may be considerably higher than in the solar system.
According to recent models for the galactic evolution of the light elements
(Vangioni-Flam et al. \cite{flam99}, Fields \& Olive \cite{fields99})
the \sixseven\ ratio reaches a maximum of about 0.3 at
a metallicity of $\feh \simeq -0.5$. At higher metallicities 
the ratio decreases due to the production of \Liseven\ in 
AGB stars, novae and supernovae of type II by the $\nu$-process
(Matteucci et al. \cite{matte95}, Woosley \& Weaver \cite{woosley95},
Vangioni-Flam et al. \cite{flam96}). Hence, it seems well justified 
to look for \Lisix\
in the metal-poor disk stars. Any detection will provide important
constraints of the chemical evolutionary models, and with a large set
of data it may also be possible to constrain the degree of
\Lisix\ depletion as a function of stellar mass and metallicity.

In the present paper we present results for the \sixseven\ ratio
for five metal-poor disk stars ranging in metallicity from $-0.8$ to
$-0.6$. Very high S/N spectra of the \Lione\ 6708\AA\ resonance
line are presented in Sect. 3 and analyzed with model atmosphere
techniques in Sect.\,4.
This has led to a rather clear detection of \Lisix\ in the two
stars with the highest masses and tight upper limits for \Lisix\
in the other stars. The consequences of these results are discussed
in Sect.\,5.

\section{Selection of stars}
The program stars were selected from the large survey
of nearby disk dwarfs by Edvardsson et al. (\cite{edv93}),
which includes accurate values of atmospheric parameters,
abundances, kinematics and ages of 189 main sequence stars
distributed in metallicity from $\feh = -1.0$ to +0.3.
As we wanted to avoid stars formed from interstellar gas greatly
enriched in \Liseven\ from AGB stars or novae,
the first condition for including a star was $\feh \la -0.5$.
Next, only stars with $\teff \ga 5900$~K were included in
order to maximize the chance of survival of some \Lisix .
Finally, the program had to be limited to a few of the brightest
stars in order to be able to reach the high S/N that is needed
to determine the lithium isotope ratio from the profile of
the 6707.8~\AA\ \Lione\ line.

The five stars observed are listed in Table 1.
The effective temperature (derived from the $b-y$ color index),
the logarithmic surface gravity, the iron abundance, and the microturbulence
velocity, are taken from Edvardsson et al. (\cite{edv93}).
Note, that none of the stars are significantly affected
by interstellar reddening according to the color excesses derived from
the $H_{\beta}$ index and $b-y$. According to the kinematical
parameters of the stars as given in Edvardsson et al. (\cite{edv93})
they belong to either the thick disk or the old thin disk.
The $\alpha$-elements, e.g. O and Mg, are somewhat enhanced in the
stars, ranging from [$\alpha$/Fe] $\simeq  0.15$ in 
\object{HR\,8181}, \object{HD\,68284} and \object{HD\,130551} to
[$\alpha$/Fe] $\simeq 0.25$ in \object{HR\,2883} and \object{HR\,3578}.

\begin{table}
\caption[ ]{List of program stars including the visual magnitude $m_V$
and atmospheric parameters adopted from Edvardsson et al. (\cite{edv93})}

\begin{tabular}{ccccccc}
\hline\noalign{\smallskip}
 HR & HD & $m_V$  & \teff & \logg  & \feh & $\xi_{micro}$      \\
    &    &        &  [K]  &        &      & [\kmprs]           \\
\noalign{\smallskip}
\hline\noalign{\smallskip}
 2883 & 59984  &  5.93  &  5976  & 4.18 & $-0.75$ & 1.7    \\
 3578 & 76932  &  5.80  &  5965  & 4.37 & $-0.82$ & 1.4    \\
 8181 & 203608 &  4.23  &  6139  & 4.34 & $-0.67$ & 1.6    \\
      & 68284  &  7.77  &  5883  & 3.96 & $-0.59$ & 1.9    \\
      & 130551 &  7.16  &  6237  & 4.25 & $-0.62$ & 1.8    \\
\noalign{\smallskip}
\hline
\end{tabular}
\end{table}

\section{Observations and data reduction}
The observations were carried out with the ESO Coud\'{e} Echelle
Spectrometer (CES) in three different periods: October 22--27, 1992,
June 6--8, 1993, and February 5--9, 1995. In Oct. 92 and Feb. 95
the CAT 1.4m telescope was applied, whereas in June 93 the 3.6m telescope
was feeding the CES through a 35m long fiber and an image slicer 
(D'Odorico et al. \cite{dodo89}). On all occasions the detector was a
front illuminated Ford Aerospace $2048 \times 2048$ CCD with 15 $\mu$m pixels. 
The CES camera has a dispersion of 1.88 \AA\,mm$^{-1}$  at 6700 \AA\ providing 
a spectral coverage of about 58 \AA\ and 0.0282 \AA\,pixel$^{-1}$. 
Different settings for the central wavelength were applied on
the various nights in order to avoid any signatures in the spectrum
resulting from a possible improper flatfielding.

\begin{figure}
\resizebox{\hsize}{!}{\includegraphics{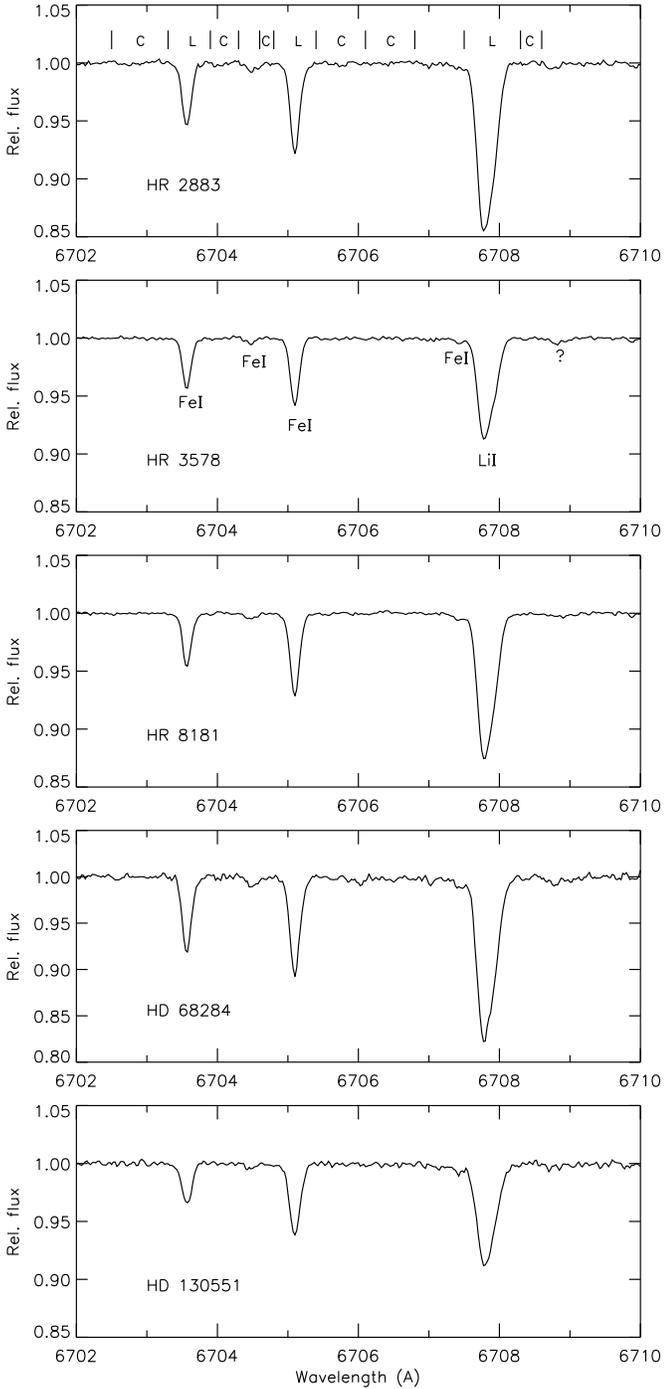}}
\caption[ ]{The observed spectra in the 6702 -- 6710~\AA\ 
region. Spectral windows used in defining the continuum
and in the $\chi^2$ analysis of the lines are marked with C and L, 
respectively. Line identifications are shown for HR~3578}
\label{fig.1}
\end{figure}

The main targets for the June 93 observations were some halo turnoff
stars, for which the results have been presented by Smith et al. (\cite
{smith98}) including a description of the reductions 
of the image slicer observations.
The Oct. 92 and Feb. 95 observations have been reduced
in much the same way. Subtraction of background, flat field correction and
extraction of spectra were performed with standard
tasks in IRAF. The wavelength calibration was based on 28 thorium lines
well distributed over the wavelength region (6676 - 6734 \AA ). A 
second-order polynomial was adopted for the dispersion solution resulting in
a typical rms deviation from the fit of 0.0015 \AA . 
Furthermore, the FWHM of the thorium lines were measured 
and found to vary by less than 4\% along the spectrum.
This near-constancy of the instrument profile is important in
connection with the determination of
instrumental and stellar line broadening
from \Feone\ lines near the \Lione\ line.
For the Oct. 92 and Feb. 95 observations the FWHM of the Th lines
corresponds to a resolution of $R = 105 \, 000$, whereas 
the June 93 image sclicer spectra have a resolution of $R = 115 \, 000$.
Finally, the spectra were normalized to an approximate
level of 1 by fitting a fifth-order cubic spline
function to the continuum, and corrected for the radial velocity
shift using the accurate wavelengths (6703.567 and 6705.102\,\AA )
given by Nave et al. (\cite{nave95}) for the two \Feone\ lines close to the 
the Li line.

\begin{table}
\caption[ ]{Number of observations and exposure times
for the program stars}

\begin{tabular}{cccc}
\hline\noalign{\smallskip}
 ID & Oct. 92  &  June 93 & Feb.95   \\
    &  CAT 1.4m &  3.6m & CAT 1.4m  \\
\noalign{\smallskip}
\hline\noalign{\smallskip}
 HR\,2883   & $4 \times 40$\,min &                   &                  \\
 HR\,3578   & $3 \times 60$\,min &  $2 \times 25$\,min & $2 \times 45$\,min \\
 HR\,8181   & $4 \times 30$\,min &  $2 \times 10$\,min & \\
 HD\,68284  &                  &                   & $8 \times 60$\,min \\
 HD\,130551 &                  &                   & $6 \times 50$\,min \\
\noalign{\smallskip}
\hline
\end{tabular}
\end{table}

Table 2 lists the the number of observations and  exposure times 
as distributed over the three periods. Note, that HR~3578 has
been observed in all three periods and HR~8181 in two of them.
The individual spectra show excellent agreement and
have therefore been co-added to obtain the final spectra
shown in Fig.~\ref{fig.1} for the 6702 -- 6710\AA\  region.
These are the spectra used for the determination of the
\sixseven\ ratio, but in the case of HR~3578 we have also 
derived the ratio separately for each observing period in order
to check for possible systematic differences.

\section{The isotope ratio \sixseven }

\subsection{The center-of-gravity wavelength of the \Lione\ line}

As discussed by e.g. Smith et al. (\cite{smith98}) the \sixseven\
ratio can be determined by two methods:  From the
center-of-gravity ($cog$) of the \Lione\ 6708\AA\ line or from a 
detailed model atmosphere synthesis of the profile.
The isotopic shift of the \Lisix\ doublet is +0.158\,\AA\ relative
to the \Liseven\ doublet. Addition of \Lisix\ therefore shifts
the \Lione\ line to longer wavelengths and  increases the FWHM.
The $cog$-method relies in principle on a very simple and straightforward
measurement, but its accuracy is limited by possible errors in the
laboratory wavelengths of the lithium line and the reference lines
needed to correct for the radial velocity
shift of the star. The errors in the laboratory wavelengths are typically
$\pm 2$\,m\AA . Differences in convective
blueshifts may well be of the same order of size (Dravins \cite{dra87}).
To this should be added the uncertainty in measuring the
$cog$-$\lambda$ for an asymmetric lithium line, which is slightly
blended by a weak \Feone\ line in the blue wing. According to our
experience the $cog$-$\lambda$ will be uncertain by about $\pm 5$\,m\AA .
This translates to a one-sigma error of $\pm 0.03$ in
\sixseven , which is inferior to what may be obtained with the
profile method. Hence, only the profile method will be used, and
the wavelength of the \Lione\ line will be considered
as a free parameter in the comparison between synthetic and observed
profiles. 

\subsection{Synthetic spectra}

\begin{figure}
\resizebox{\hsize}{!}{\includegraphics{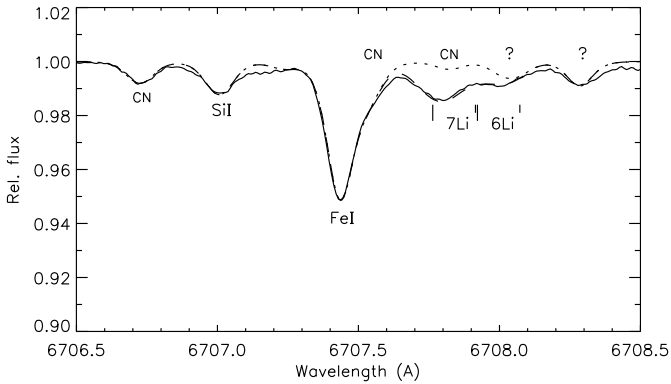}}
\caption[ ]{The solar flux spectrum around the \Lione\ line.
The full drawn line is the observed spectrum from the
Solar Flux Atlas of Kurucz et al. (\cite{kurucz84}).
The dashed line is the synthetic spectrum computed for the
model atmosphere of the Sun. The dotted line is the synthetic spectrum
without the \Lione\ line}
\label{fig.2}
\end{figure}
 
The synthetic spectra have been obtained with the Uppsala
Synthetic Spectrum Package, which computes the flux in a given
wavelength region for an OSMARC model atmosphere of the star
(Edvardsson et al. \cite{edv93}). LTE is assumed and
thermal and microturbulent line broadening as well as
pressure broadening is included. The resulting spectrum is
folded with a line broadening function that can be either
Gaussian, radial-tangential (Gray \cite{gray78}), rotational
(Gray \cite{gray92}) or any
combination of these functions. As a check, one star (\object{HD\,68284})
has also been analyzed with a Kurucz ATLAS9 model atmosphere and the
SYNTHE code (Kurucz \cite{kurucz93}).

Wavelengths and $gf$-values for the \Lisix\ and \Liseven\ components
of the \Lione\ doublet are taken from Table 3 of Smith et al. 
(\cite{smith98}). 
As discussed by Smith et al. these values are not a source of significant
errors in connection with determinations of the isotopic abundances of Li.

In the case of halo turnoff stars with $\feh < -1.5$ there is no
significant blending of the \Lione\ line by other lines, but already
at $\feh \simeq -1.0$ one has to worry about contributions 
from other absorption lines. To get more detailed information about
line blending we therefore started by synthesizing the solar spectrum, which is
well suited for studying this problem because of the weakness of
the \Lione\ line. Fig.~\ref{fig.2}  shows the solar flux spectrum from the
atlas of  Kurucz et al. (\cite{kurucz84}) compared to
a synthetic spectrum based on the OSMARC solar model and a list of 
CN and metal lines from M\"{u}ller et al. (\cite{muller75}),
who obtained log\,$\epsilon$(Li) = 1.0 and $\sixseven \simeq 0.0$
from a synthesis analysis of the solar \Lione\ feature.
We adopted these values and adjusted the gf-values of the other lines
to get the best possible fit of the synthetic spectrum 
(broadened by a radial-tangential function with a FWHM = 2.5\,\kmprs )
to the Solar Flux Atlas.
As seen from Fig.~\ref{fig.2} the fit is quite satisfactory. The
main problem is the two unidentified weak lines at 6708.02 and
6708.28\,\AA , which have equivalent widths of 0.6 and 1.1\,m\AA , 
respectively. The first line has nearly the same wavelength as the weak 
component of the \Lisix\ doublet and therefore makes the determination
of \sixseven\ very tricky for solar-type metallicities, whereas the
unidentified line at 6708.28\,\AA\ is further away and only
affects the determination of the lithium isotope ratio marginally.

Assuming that the two unidentified lines are neutral metal lines
they will have equivalent widths less than 0.15 and
0.3\,m\AA , respectively, in 
solar-type stars with $\feh \la -0.6$.
This is too small to have any significant effect
on the determination of the lithium isotope ratio. If the lines
are ionized metal lines they will be somewhat stronger
and affect the determination of \sixseven\ marginally at the 
1\% level. The probability for lines in solar-type spectra being
from ionized metals instead of neutral is, however, small, and we
have therefore chosen to exclude the two unidentified lines from our
model atmosphere synthesis of the metal-poor disk stars. 

The CN lines seen in Fig.~\ref{fig.2} play no r\^{o}le in the spectra
of the program stars mainly because both C and N  scales almost
linearly with Fe and partly because the stars have somewhat
higher effective temperatures than the Sun. Hence, there is no reason to
include these lines. The only line that has a significant effect on
the synthesis
of the lithium line is the \Feone\ line at 6707.43\,\AA . This line
was included with the $gf$-value derived from the fit to the solar
spectrum.
 
As discussed in Sect. 4.3 the \Feone\ lines at
6703.6 and 6705.1\,\AA\ have been used to determine the
instrumental and stellar
atmospheric line broadening. The two lines are blended by several
faint CN lines in the solar spectrum as seen from Fig.\,2 of 
Brault \& M\"{u}ller (\cite{brault75}), but the synthesis of the 
solar and stellar spectra shows that these lines disappear
beyond detection in the program stars. The same technique has been
used to define a number of spectral windows practically 
free of lines in the program stars. These regions are marked  by `C' in 
Fig.~\ref{fig.1}, and are used to set the continuum and to
estimate the S/N in connection with the chi-square analysis of the lines. 

\subsection{$\chi{^2}$ analysis}
The synthetic spectra were computed for model atmospheres
with the parameters given in Table 1.
Line broadening due to macroturbulence and
rotation was determined from the two \Feone\ lines at
6703.6 and 6705.1\,\AA\ with $\chi_{exc} = 2.76$ and 4.61\,eV, respectively.
The synthetic spectrum was first folded with a Gaussian function 
representing the instrumental profile and then with either a Gaussian,
a radial-tangential or a rotational profile. Various combinations of
these profiles was also tried, but in no cases the fit to the iron
lines was better than that obtained with a Gaussian function.
The final analysis was therefore carried out with a single Gaussian
representing the combined effect of instrumental and stellar atmospheric
broadening.

\begin{figure}
\resizebox{\hsize}{!}{\includegraphics{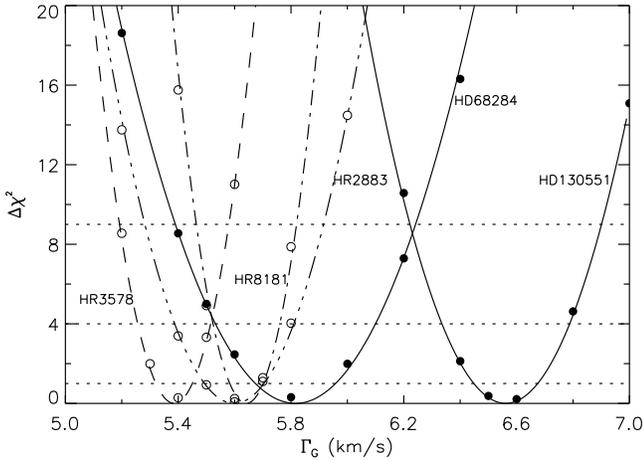}}
\caption[]{Variation of the $\chi^{2}$ of the fit to the 
\Feone\ 6705.1\,\AA\ line as a function of the FWHM of
the Gaussian broadening function}  
\label{fig.3}
\end{figure}

The FWHM of the Gaussian broadening 
profile, $\Gamma_{G}$, has been determined with the following procedure.  
First the
continuum is set from the two windows on each side of a \Feone\ line and at
the same time the standard deviation, $\sigma\,=\,(S/N)^{-1}$
of the spectrum is estimated. Then the chi-square
function is computed:
\begin{equation}
\chi^{2} = \Sigma[ \frac {(O_i - S_i)^{2}}  {\sigma^{2}} ]
\end{equation}
where O$_{i}$ is the observed spectral point and S$_{i}$ is the
synthesis. The summation is performed over the spectral interval
marked by `L' in Fig.\,1 corresponding to N\,=\,25 data\-points.
In addition to $\Gamma_{G}$, the equivalent width
and the exact wavelength of the \Feone\ line are
considered as free parameters. $\Gamma_{G}$ is varied in steps
of 0.1 or 0.2~\kmprs\ and the other two parameters are
optimized for each value of $\Gamma_{G}$ to find the lowest $\chi^{2}$.
This results in a parabolic variation of $\chi^{2}$ as shown in
Fig.~\ref{fig.3}. The most probable value of $\Gamma_{G}$ corresponds
to the minimum of $\chi^{2}$, and $\Delta \chi^{2}$\,=\,1, 4 and 9
correspond to the 1-, 2-, and 3-$\sigma$ confidence limits
of determining $\Gamma_{G}$ alone (Bevington \& Robinson \cite{bev92}).

Table 3 summarizes the results of the $\chi^{2}$ fitting of
the two \Feone\ lines. Note, that the reduced
chi-square ($\chi^{2}_{red} = \chi^{2} / \nu$,
where $\nu =22$ is the number of degrees of
freedom in the fit) is satisfactorily close to 1.
Furthermore, the values of
$\Gamma_{G}$ from the two \Feone\ lines agree rather well,
although there is a tendency that the 6705.1\,\AA\ line gives slightly 
higher values of $\Gamma_{G}$ than the 6703.6\,\AA\ line.

\begin{table}
\caption[]{Results from the $\chi^{2}$ analysis of the \Feone\ lines
at 6703.6 and 6705.1\,\AA .  $\Gamma_{G}\,(6703)$
and $\Gamma_{G}\,(6705)$ are the FWHM of the Gaussian broadening function
(instrumental + rotation + macroturbulence) applied to the synthetic lines.
The errors given are the formal one-sigma errors resulting from the
$\chi^{2}$ analysis}

\begin{tabular}{cccccc}
\hline\noalign{\smallskip}
 ID &  S/N & $\Gamma_{G}\,(6703) $  & $\chi^{2}_{red}$ &
                     $\Gamma_{G}\,(6705) $  & $\chi^{2}_{red}$   \\
           
    &      &          [\kmprs]      &                  &    
                      [\kmprs]      &                       \\
\noalign{\smallskip}
\hline\noalign{\smallskip}
 HR\,2883    &  700 & 5.70 $\pm .17$ & 0.78 & 5.60 $\pm .11$ & 1.36  \\
 HR\,3578    & 1250 & 5.35 $\pm .09$ & 1.07 & 5.40 $\pm .06$ & 1.05  \\
 HR\,8181    & 1370 & 5.25 $\pm .08$ & 1.17 & 5.65 $\pm .05$ & 1.22  \\
 HD\,68284   &  400 & 5.40 $\pm .20$ & 0.96 & 5.80 $\pm .14$ & 0.93  \\
 HD\,130551  &  830 & 6.30 $\pm .15$ & 1.05 & 6.55 $\pm .10$ & 0.96  \\
\noalign{\smallskip}
\hline
\end{tabular}
\end{table}

\begin{figure}
\resizebox{\hsize}{!}{\includegraphics{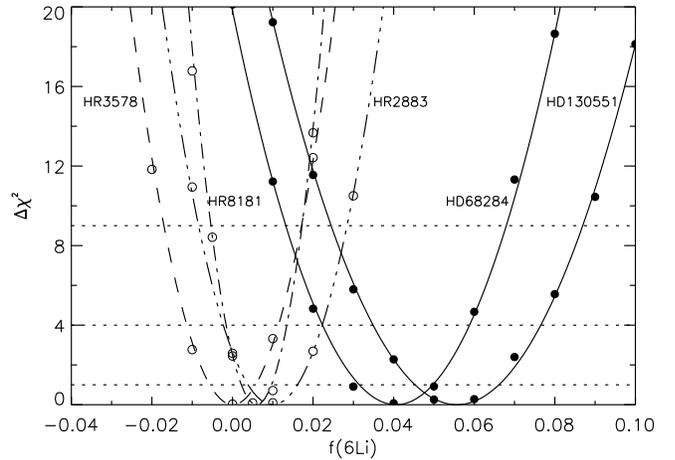}}
\caption[]{Variation of the $\chi^{2}$ of the fit to the 
\Lione\ 6707.8\,\AA\ line as a function of the relative abundance
of \Lisix }
\label{fig.4}
\end{figure}

Adopting a weighted average of $\Gamma_{G}$ from Table 3
the lithium isotope ratio is determined from a $\chi^{2}$
analysis of the fit between the computed and observed profile
of the \Lione\ line. The free parameters in the fit are the total
lithium abundance of the star, log$\epsilon$(Li), and the \Lisix\ fraction,
$f(\Lisix) = N(\Lisix)/N$(Li). Furthermore, a
wavelength shift, $\Delta \lambda$, of the Li line relative to the
wavelengths given in Table 3 of Smith et al. (\cite{smith98}) is allowed.
The continuum and the S/N are determined from the two adjacent windows shown in 
Fig.~\ref{fig.1} (resulting in nearly the same S/N values as given in
Table 3) and the $\chi^{2}$ analysis is carried out over the
region marked by `L', i.e. over N=33 datapoints. Note, that the weak
\Feone\ line at 6707.43\,\AA\ has only a small effect on
the flux in this line region.

\begin{figure}
\resizebox{\hsize}{!}{\includegraphics{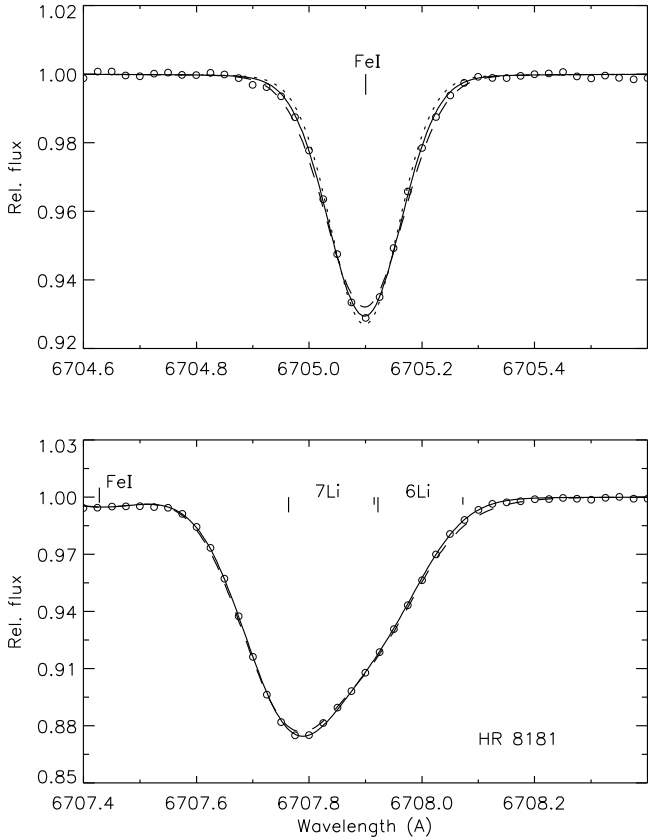}}
\caption[]{The model atmosphere synthesis of the \Feone\ 6705.1\,\AA\
and the \Lione\ 6707.8\,\AA\ line in the spectrum of \object{HR\,8181}.
The datapoints are shown with open circles. In the upper figure the
full drawn line corresponds to a Gaussian broadening parameter of
$\Gamma_{G}=5.6\,\kmprs$, whereas the dotted and dashed lines correspond to
$\Gamma_{G}=5.0$ and 6.2\,\kmprs , respectively. In the lower figure
$\Gamma_{G}=5.5\,\kmprs$ has been applied. The full drawn line corresponds to
$f(\Lisix) = 0.0$ and the dashed line to $f(\Lisix) = 0.05$. Note,
that when $\Gamma_{G}$ and $f(\Lisix)$ are varied the other free
parameters in the fits, the wavelengths and the equivalent widths
of the lines, have been optimized to get the best possible fits}
\label{fig.5}
\end{figure}

\begin{figure}
\resizebox{\hsize}{!}{\includegraphics{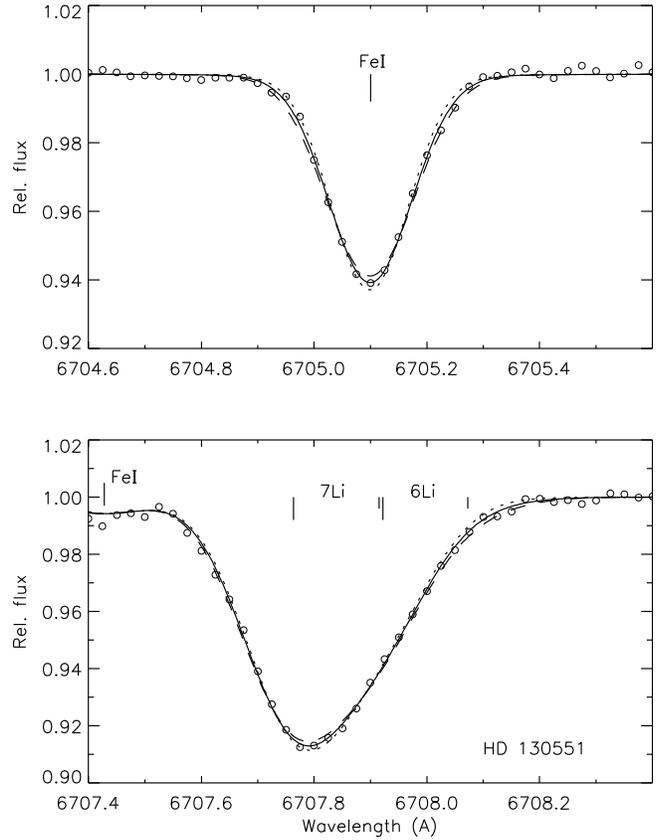}}
\caption[]{Same as Fig.~\ref{fig.4} for \object{HD\,130551}.
In the upper figure the
full drawn line corresponds to a Gaussian broadening parameter of
$\Gamma_{G}=6.6\,\kmprs$, and the dotted and dashed lines to
$\Gamma_{G}=6.0$ and 7.2\,\kmprs , respectively. In the lower figure
$\Gamma_{G}=6.6\,\kmprs$ has been applied. Here the full drawn line
corresponds to
$f(\Lisix) = 0.06$ and the dotted and dashed lines to $f(\Lisix) = 0.00$
and 0.10, respectively}
\label{fig.6}
\end{figure}

The relative \Lisix\ abundance is varied in steps of 0.01 and 
the other two parameters, log$\epsilon$(Li) and $\Delta \lambda$,
are optimized for each value of $f(\Lisix)$ to find the lowest $\chi^{2}$.
In order to study the behaviour of the $\chi^{2}$ function around
$f(\Lisix) = 0.00$ `negative' values of $f(\Lisix)$ have been simulated
by subtracting the line absorption coefficient due to \Lisix\ from
the continuous absorption coefficient instead of adding it.

\begin{figure}
\resizebox{\hsize}{!}{\includegraphics{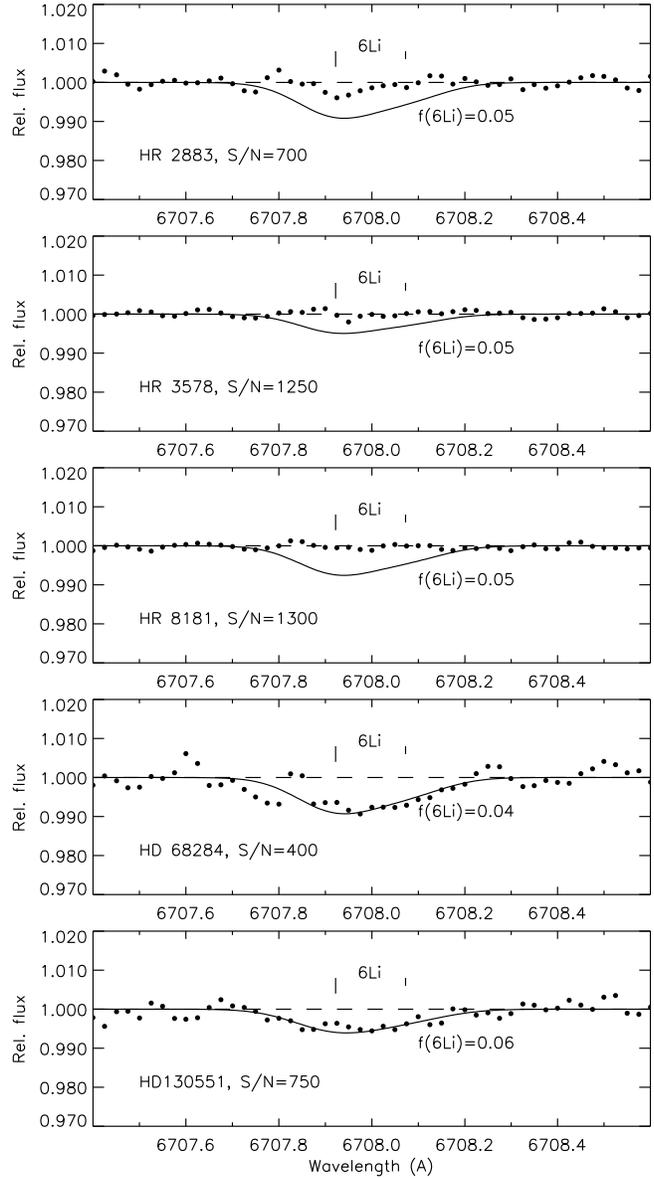}}
\caption[]{The residuals of the observations 
after subtraction of the \Liseven\ and \Feone\ 6707.43~\AA\ part of the
synthesis of the \Lione\ line. For comparison the synthesis of
the \Lisix\ doublet is shown with a full drawn line}
\label{fig.7}
\end{figure}

The results of the $\chi^{2}$ analysis are summarized in Table 4 and
the variation of $\chi^{2}$ with $f(\Lisix)$ is shown in
Fig.~\ref{fig.4}. As seen,
the three `HR' stars have $f(\Lisix)$ close to zero, whereas \Lisix\
has been detected in the two `HD' stars at a high confidence level.
As a further illustration of this result Figs.~\ref{fig.5} and \ref{fig.6}
show the fits to the \Feone\ and the \Lione\ for two stars -- one
with $f(\Lisix) \simeq 0.0$ and the
other one with $f(\Lisix) \simeq 0.06$ --
and Fig.~\ref{fig.7} shows a plot of the residuals in the
observations after subtracting the \Liseven\ and \Feone\ 
part of the synthesis. Although there is a disturbing periodic noise
in the residuals with an amplitude of 0.1 to 0.2\%
in addition to the shot noise (the reason for which remains unexplained)
a clear residual absorption at the wavelength of the 
\Lisix\ doublet is seen in the spectra
of \object{HD\,68284} and \object{HD\,130551}.
Note, that although the S/N of the spectrum of \object{HD\,68284} is
inferior to that of \object{HD\,130551}, the error of $f(\Lisix)$
is nearly the same for the two stars, because the Li line
is about a factor of two stronger in the spectrum of
\object{HD\,68284} than in the case of \object{HD\,130551}.

In order to check that there are no significant systematic differences
between the results obtained for the three observing periods 
the individual spectra of \object{HR\,3578} were analyzed separately 
including the determination of $\Gamma_G$.
The $\chi^{2}$ analysis gave the following values: 
$f(\Lisix) = -0.018 \pm 0.012$, $ +0.008 \pm 0.010$, and
$-0.006 \pm 0.014$ for the Oct. 92, June 93 and Feb. 95 spectra.
Within the errors these values agree satisfactorily with the
value ($f(\Lisix) = 0.000 \pm 0.006$) derived for the averaged spectrum.

\begin{table}
\caption[]{Results from the $\chi^{2}$ analysis of the \Lione\ line.
$\Gamma_{G}$ is the weighted average
of the FWHM of the Gaussian broadening function determined
from the \Feone\ lines at 6703.6 and 6705.1\,\AA\ (Table 3). 
log$\epsilon$(Li) is the total Li abundance and $f(\Lisix)$
is the relative abundance of \Lisix . The errors given are the
formal one-sigma errors resulting from the $\chi^2$ analysis.
The last column gives the reduced chi-square of the fit}

\begin{tabular}{ccccc}
\hline\noalign{\smallskip}
 ID &  $\Gamma_{G}$ & log$\epsilon$(Li) & $f(\Lisix)$ 
 & $\chi^{2}_{red}$   \\
    & [\kmprs ]     &                   &           &         \\
\noalign{\smallskip}
\hline\noalign{\smallskip}
 HR\,2883    & 5.7          & 2.34 & 0.010 $\pm .007$ & 0.99  \\
 HR\,3578    & 5.4          & 2.05 & 0.000 $\pm .006$ & 1.03  \\
 HR\,8181    & 5.5          & 2.38 & 0.004 $\pm .004$ & 0.89  \\
 HD\,68284   & 5.6          & 2.35 & 0.041 $\pm .010$ & 1.09  \\
 HD\,130551  & 6.5          & 2.30 & 0.056 $\pm .011$ & 1.22  \\
\noalign{\smallskip}
\hline
\end{tabular}
\end{table}

\subsection{Systematic errors of \sixseven }
The errors of $f(\Lisix)$ given in Table 4 are purely statistical.
In addition we must consider possible systematic errors resulting
from approximations in the model atmospheres and the spectrum
synthesis of the \Feone\ and \Lione\ lines.

First we note that the exact values of the atmospheric parameters
of the models are not critical for the lithium isotope ratio
derived. An increase of \teff\ 
with +100\,K changes  log$\epsilon$(Li) with +0.07\,dex but 
$f(\Lisix)$ is practically unchanged. Changes in the gravity within
reasonable limits have completely negligible effects. A decrease of the
microturbulence by say 0.4\,\kmprs\
is compensated by a slight increase of $\Gamma_G$
and the change of $f(\Lisix)$ is less than 0.003.
A more significant, but still rather small change, results from the
use of Kurucz's model atmospheres (Kurucz \cite{kurucz93})
instead of the OSMARC models. As an example a model atmosphere
with the parameters of \object{HD\,68284} was interpolated
between the closest models in the Kurucz ATLAS9 grid (re-computed with
the ``approximate" overshooting option switched off)
and the SYNTHE code was used to compute profiles
of the \Feone\ and \Lione\ lines. This resulted in an increase of
$\Gamma_G$ from 5.6 to 5.9\,\kmprs\ and a change of $f(\Lisix)$
from 0.041 to 0.033. Hence, the class of plane-parallel 
model atmospheres adopted has some effect on the value of the
lithium isotope ratio derived, but not large enough to question
the detection of \Lisix\ in \object{HD\,68284} and \object{HD\,130551}.

A more critical problem is whether classical plane-parallel
model atmospheres are good enough for a determination
of $f(\Lisix)$ with an accuracy of say $\pm 0.01$. It is well known
that convective motions in real atmospheres produce slightly asymmetric
line profiles, i.e. curved bisectors with a shape that depends on the
depth of line formation (e.g. Dravins \cite{dra87}). The question is
therefore if the atmospheric velocity broadening 
determined from the two \Feone\ lines is also valid for the \Lione\
line. Due to the low excitation potential of the lithium resonance line
it is probably formed somewhat higher in the atmosphere 
than the iron lines. In the case of \object{HD\,130551} a drastic
increase of $\Gamma_G$ to 7.5\,\kmprs\ instead of the value
derived from the \Feone\ lines,
(6.5\,\kmprs ) decreases $f(\Lisix)$ to about
zero. One could also imagine that the \Lione\ line in \object{HD\,68284}
and \object{HD\,130551} has a red asymmetry that mimics the \Lisix\
doublet although no such asymmetry is seen in the profiles of the
\Feone\ lines. However, such possible effects have to occur for 
\object{HD\,68284} and \object{HD\,130551} only, because we can 
not allow a reduction of $f(\Lisix)$ for the other three stars
which have $f(\Lisix) \simeq 0.00$, when the atmospheric
velocity broadening of the \Feone\ lines is adopted.

As seen from Table 1 there are indeed significant differences in the
atmospheric parameters of the two `HD' stars and the three `HR' stars
that could induce some differential effects in the broadening of
the \Feone\ and \Lione\ lines. \object{HD\,68284} has a lower gravity
and \object{HD\,130551} has a higher \teff\ than the other
three stars. A hint that there may be systematic differences in the
convective pattern between the two groups of stars
comes from the small changes of the laboratory wavelengths needed
to optimize the $\chi^2$ fit of the \Feone\ and \Lione\ lines
(see Table 5). The apparent heliocentric radial velocities 
(including gravitational redshift and convective blueshift) is
determined from the two \Feone\ lines. Hence, the sum of the
wavelength shifts for these two lines is zero by definition.
As seen, the wavelength shift of the \Lione\ line is always
positive. The average value is 4.4\,m\AA\ (corresponding to a
redshift of 0.20\,\kmprs\ of the \Lione\ line relative to
the \Feone\ lines) and the rms scatter is
3.4\,m\AA . This is more than the expected error of the laboratory
wavelengths. In particular, we note that the redshift for 
\object{HD\,68284} and \object{HD\,130551} are lower
than the redshift for the other three stars. Although this
could be accidental, it is a warning that there may be
differential effects  in the convective line broadening.
Clearly, this problem should be
further studied by applying recently constructed inhomogeneous
3D hydrodynamical model atmospheres (Asplund et al. \cite{asp99})
in the analysis of the \Lione\ line.

\begin{table}
\caption[]{Radial velocities of the stars as derived from the
accurate wavelengths of the \Feone\ lines  (6703.567
and 6705.102\,\AA ) measured by  Nave et al. (\cite{nave95}).
The wavelengths shifts  given are
obtained from the $\chi^2$ fits of the synthetic spectra to
the individual lines}

\begin{tabular}{ccccc}
\hline\noalign{\smallskip}
 ID & RV &  $\Delta \lambda(6703)$ & $\Delta \lambda(6705)$ &
       $\Delta \lambda(\Lione)$ \\
    & [\kmprs] & [m\AA] & [m\AA] & [m\AA]  \\
\noalign{\smallskip}
\hline\noalign{\smallskip}
 HR\,2883    &  +54.8 & $-0.1$  &  +0.0  &  4.3 \\
 HR\,3578    &  +119.4&  +1.3   & $-1.4$ &  9.1      \\
 HR\,8181    & $-28.8$ &  +2.6   & $-2.5$ &  6.7      \\
 HD\,68284   &  +62.8 &  +1.3   & $-1.4$ &  1.8      \\
 HD\,130551  &  +33.7 &  +1.4   & $-1.4$ &  0.1      \\
\noalign{\smallskip}
\hline
\end{tabular}
\end{table}

\section{Discussion}

\subsection{Stellar masses}
According to standard stellar models the depletion of lithium
is a strong function of stellar mass (Pinsonneault et al. \cite{pin92}).
Hence, it is of considerable interest to determine the masses of the stars.
This can be done by comparing \teff\ and absolute magnitude, $M_V$,
with mass tracks from stellar evolution calculations.

\begin{figure}
\resizebox{\hsize}{!}{\includegraphics{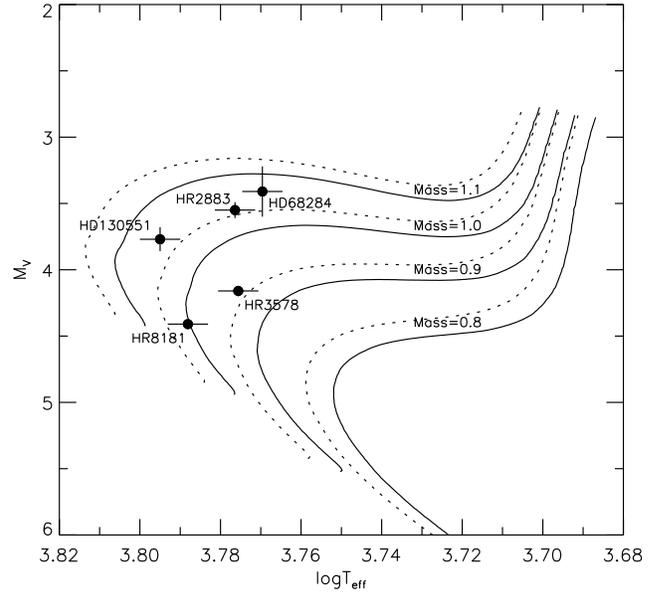}}
\caption[]{Position of the stars in the \logteff -$M_V$ diagram
compared with mass tracks from VandenBerg et al. (\cite{van99}).
Masses are given in units of the solar mass.
The full drawn lines refer to $\feh = -0.61$ (equal to the
metallicity of \object{HD\,68284} and \object{HD\,130551})
and the dotted lines to $\feh = -0.71$ (approximately equal
to \feh\ of the other stars).
Both sets of tracks have [$\alpha$/Fe] = 0.3.
The error bars in the x-direction correspond to $\sigma (\teff) = \pm 70$\,K
and those in the y-direction to the errors of the Hipparcos parallaxes}
\label{fig.8}
\end{figure}

Using the apparent magnitudes given in Table 1 and parallaxes from
The Hipparcos and Tycho Catalogues (ESA \cite{esa97}) the absolute
magnitudes are calculated (Table 6), and the stars are
plotted in the \logteff -$M_V$ diagram (Fig.~\ref{fig.8}). The mass tracks
shown are from the new, $\alpha$-element enhanced, evolutionary models of
VandenBerg et al. (\cite{van99}).
Interpolation between the mass tracks (taking into account their dependence
on \feh ) leads to the masses given in Table 6, and from the corresponding
isochrones the stellar ages given are obtained.

\begin{table}
\caption[]{Absolute magnitudes computed from $m_V$ and Hipparcos
parallaxes and masses and ages derived from stellar
evolutionary tracks and isochrones of VandenBerg et al. (\cite{van99})}

\begin{tabular}{cccr}
\hline\noalign{\smallskip}
 ID & $M_V$ &  ${\cal M}/{\cal M}_{\sun}$ & Age\,[Gyr]  \\
\noalign{\smallskip}
\hline\noalign{\smallskip}
 HR\,2883    & 3.55 $\pm .06$ & 1.02 $\pm .02$ &  7.6 $\pm 0.6$      \\
 HR\,3578    & 4.16 $\pm .04$ & 0.88 $\pm .02$ & 11.7 $\pm 1.2$      \\
 HR\,8181    & 4.41 $\pm .01$ & 0.96 $\pm .02$ &  6.3 $\pm 2.0$      \\
 HD\,68284   & 3.41 $\pm .19$ & 1.07 $\pm .04$ &  6.8 $\pm 1.4$      \\
 HD\,130551  & 3.77 $\pm .09$ & 1.06 $\pm .02$ &  6.0 $\pm 0.9$      \\
\noalign{\smallskip}
\hline
\end{tabular}
\end{table}

The errors of the masses and ages given in Table 6 are standard errors
corresponding to the adopted errors of \teff\ and $M_V$. Additional
errors may be present due to inadequate stellar models and uncertainties
in the calibration of \teff\ and the bolometric correction. Such errors
are, however, more systematic and are expected to affect all stars
with about the same amount. Hence, we conclude from Table 6 that
the two stars for which \Lisix\
has been detected (\object{HD\,68284} and \object{HD\,130551}) have
significantly higher masses than the three stars with no
\Lisix\ present in their atmospheres. This makes sense, because the
depth of the convection zone of a star on the main sequence 
decreases rapidly as a function of increasing mass. Hence, according
to standard stellar models without mixing, the depletion of \Lisix\ is
less severe in the more massive stars. In this connection
we note that although \object{HD\,68284} is already on the subgiant branch
and the coolest of the stars, it has spent most of its life as a
main sequence star at  $\teff \simeq 6300$~K .

\subsection{Galactic evolution and stellar depletion of \Lisix }

Interpretation of the novel result of this paper - the detection and
quantitative measurement of the $^6$Li abundance in two
old metal-poor disk stars - is
contingent on two factors: (i) the
expected evolution of the interstellar $^6$Li abundance with metallicity,
and (ii) the depletion of the stellar $^6$Li abundance by the
convective mixing that occurs in the pre-main sequence phase, and
the additional depletion occurring on the main sequence. 

As is all too well known, prediction of Li depletion by main
sequence stars and subgiants is an imprecise art. Standard models
by Pinsonneault et al. (\cite{pin92}) predict loss of lithium in the pre-main
sequence phase and no subsequent loss for stars of the mass of our
quintet. Depletions for masses of up to 0.85${\cal M}_{\sun}$ and metallicities
corresponding to [Fe/H] = $-2.6$ and $-1.6$ are computed by 
Pinsonneault et al.  For their 0.85${\cal M}_{\sun}$
model, the predicted $^7$Li-depletions are negligible and the
$^6$Li-depletions are 0.3 dex at [Fe/H] = $-1.6$ and by extrapolation
less than 0.05 dex at [Fe/H] = $-2.6$. Extrapolation to [Fe/H] $\simeq$
$-0.7$ is uncertain but these depletions decrease with increasing
mass such that our stars might be anticipated to have lost
little, if any, $^6$Li.
Cayrel et al. (\cite{cayrel99b}) report calculations
that essentially confirm the above pre-main sequence depletions 
but predict a substantial continuing depletion of $^6$Li on the main
sequence.
At ${\cal M}=0.85{\cal M}_{\sun}$ and [Fe/H] = $-1.5$, a total $^6$Li depletion of about
0.7 dex is predicted
in contrast to the 0.3 dex expected by Pinsonneault et al. Consideration
of non-standard physics, especially rotationally-induced mixing will
result in  likely larger and as yet more uncertain depletions -- see,
for example, the 
state of the art calculations by Pinsonneault et al. In summary, 
depletion of $^6$Li is to be expected but, at present, the
magnitude of this depletion is uncertain with even standard calculations
unavailable for the mass and metallicity of our old disk stars.

Encouraged by recent observations of $^6$Li, Be, and B 
several predictions about the galactic chemical evolution of Li, Be, and B
have appeared.
Behind such predictions are assumptions about the nucleosynthetic processes
of Li, Be, and B manufacture that demand assumptions about
the early Galaxy, especially about the cosmic rays that permeated the
halo and then the disk. Qualitatively, the key nucleosynthetic
processes are known: (i) the Big Bang provided only the $^7$Li (in
addition to H, $^2$H, $^3$He, and $^4$He) that is widely
considered to account for the observed  Li
abundance of the warm halo stars, the so-called Spite plateau; (ii) interactions
between standard GCR and ISM and/or interactions between fast C,N,O nuclei
from superbubbles with H or He in the ISM provide
Li, Be, and B by spallation processes (e.g., O + p $\rightarrow$ Be) and Li
through the fusion process $\alpha + \alpha \rightarrow ^6$Li and $^7$Li;
(iii) neutrino-induced spallation processes in Type II supernovae that
may provide $^7$Li and $^{11}$B.  

A key facet of this suite of processes is that beryllium with $^9$Be as the
single
stable isotope is produced solely by spallation of C,N,O in flight or at rest.
Hence, observed beryllium abundances may  be used to calibrate the yields
of cosmic ray spallation.
This is especially useful now that there
are  extensive measurements of the Be abundance in disk and halo
dwarf stars. The relative
yields of light nuclides, for example $^6$Li to $^9$Be, are essentially
independent  of the cosmic ray spectrum unless there is a large excess of
low energy cosmic rays ($E < 30$ MeV nucleon$^{-1}$) with respect to
higher energy particles. This happy circumstance  arises because
above the similar threshold energies for the different processes 
(e.g., $^9$Be from
$p$ + O and $^{10}$B also from $p$ + O), the spallation cross-sections are
almost energy independent. That ratios of yields are independent of the form
of the (high) energy spectrum was well illustrated by Ramaty et al. 
(\cite{ramaty96}).
At energies around the threshold energies for the various
processes, the relative yields are energy and composition dependent. 
Moreover, Li production occurs also through $\alpha + \alpha$ 
fusion reactions that do not synthesize Be and B.

The  B/Be ratio of halo stars is consistent within measurement
uncertainty  with production by spallation:
Duncan et al. (\cite{duncan97}) estimated B/Be = 15 $\pm$ 3
and Garc\'{\i}a L\'{o}pez et al. (\cite{garcia98}) from a similar
dataset of HST spectra
found B/Be = 17 $\pm$ 10. Relativistic cosmic rays and the suite of
(p,$\alpha$) on (C,N,O) processes are predicted to give
B/Be $\simeq 14$  --
see Ramaty et al. (\cite{ramaty96}) for predicted B/Be ratios as a function
of cosmic ray energy and composition. It has long been known that
spallation by relativistic cosmic rays is inadequate to account for
the solar system's  $^{11}$B/$^{10}$B ratio which at 4.05 exceeds the
prediction of about 2.5. Low energy spallation or a contribution from Type II
supernovae are needed to resolve this discrepancy. 

\begin{figure}
\resizebox{\hsize}{!}{\includegraphics{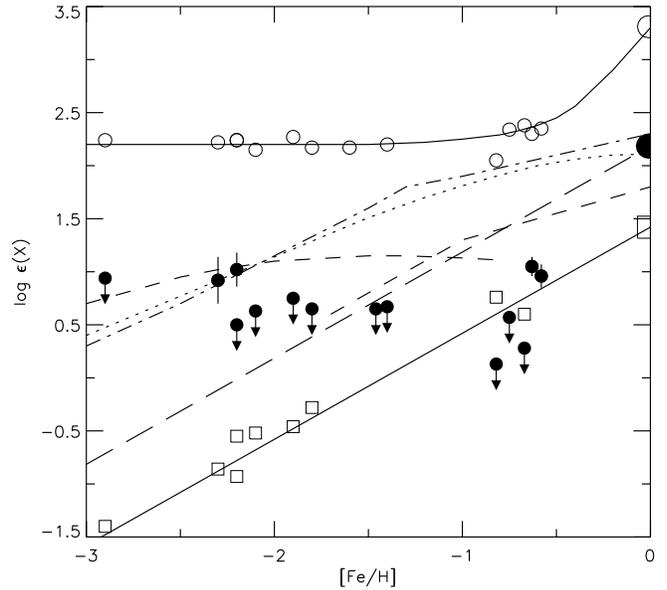}}
\caption[]{Abundances of lithium and beryllium as a function of \feh\ for
9 halo stars from Smith et al. (\cite{smith98}) and 5 disk stars from
the present paper. Open circles indicate the total Li abundance
and filled circles the \Lisix\ abundance or an upper limit. Open
squares refer to the Be abundance adopted from 
Gilmore et al. (\cite{gil92}) and Boesgaard et al. (\cite{boes99b}). 
The big symbols indicate meteoritic abundances from
Anders \& Grevesse (\cite{anders89}). The upper full drawn line is a
fit to the `Spite plateau' of lithium abundances for $\feh < -1.5$ and to the
upper envelope of the Li abundance distribution for disk stars
(see fig. 7 of Lambert et al. \cite{lam91}). The lower full drawn line is
a linear fit to the beryllium abundances with a slope of one, and the
long-dashed line shows the corresponding relation for
\Lisix\ if \Lisix /Be = 5.8
as found in meteorites and as predicted from spallation of CNO nuclei by
high energy cosmic rays.
The dotted line represents the evolution of \Lisix\ in the model of
Fields \& Olive (\cite{fields99}),
the dashed-dotted line the model of Vangioni-Flam et al.  (\cite{flam99})
and the short-dashed lines refer to the model of Yoshii et al. (\cite{yoshii97})
for the halo and the disk, respectively}
\label{fig.9}
\end{figure}

These uncertainties aside, the Be observations are a reasonably firm basis
from which to predict the $^6$Li abundances provided by 
spallation. Smith et al. (\cite{smith98}) discussed the
prediction of $^6$Li abundances from observed Be abundances
- see the long-dashed line in Fig.~\ref{fig.9}, where beryllium abundances
are taken from Gilmore et al. (\cite{gil92}) and Boesgaard et al.
(\cite{boes99b}). Predicted $^6$Li abundances are
about a factor of 10 less than the observed $^6$Li abundances in the
halo stars \object{HD\,84937} and \object{BD\,+26$^{\circ}$3578}
but exceed the $^6$Li abundances
reported here for the disk stars \object{HD\,68284} and \object{HD\,130551}
by about a factor of 3. 
Since $^6$Li has almost certainly been depleted during pre-main sequence
evolution and possibly during residence on the main sequence, the
initial or interstellar $^6$Li abundance for the halo stars was
higher than now observed.
The required additional $^6$Li is probably primarily
a product of cosmic ray  $\alpha + \alpha$ fusion production.

Predictions of the growth of $^6$Li in the Galaxy  made 
recently by Fields \& Olive (\cite{fields99}),
Vangioni-Flam et al. (\cite{flam99}) and Yoshii et al. (\cite{yoshii97}) are
shown in Fig.~\ref{fig.9}. These call on the same
production processes but in different proportions.

Fields \& Olive discuss what they term the standard picture of
galactic cosmic ray nucleosynthesis in  a model galaxy. The key
assumptions are that 
the cosmic rays always  had the composition of the ambient
interstellar gas (i.e., they were very CNO-poor early in the life of the
Galaxy), the energy spectrum of the cosmic rays was that measured
for contemporary cosmic rays in the solar neighborhood (i.e., relativistic
energies are dominant), 
and  the cosmic ray flux has been proportional
to the local supernova rate, and scaled so that solar abundances
of $^6$Li, B, and $^{10}$B    are reproduced.  These assumptions
with a simple chemical evolution code (Fields \& Olive report
results for the canonical closed box) lead to the predicted run
of the $^6$Li abundances with [Fe/H] where iron is a product of
stellar nucleosynthesis with yields from Woosley \& Weaver (\cite{woosley95})
and a standard initial mass function. The key novel ingredient in 
the otherwise familiar calculation  is the incorporation of recent
measurements of the oxygen abundance in halo stars that indicate [O/Fe]
increasing with decreasing [Fe/H] (Israelian et al. \cite{israel98};
Boesgaard et al. \cite{boes99a}). A higher O abundance increases
the yields of spallation products. (A `fudge' is needed as the O/Fe
from these recent observations is considerably higher than predicted
for Type II supernovae).  The predicted $^6$Li vs [Fe/H] relation
is shown by the dotted line in Fig.~\ref{fig.9}. A large part of the increase
in the $^6$Li prediction at low [Fe/H] over the simple expectation 
from spallation is due to the inclusion of the
$\alpha + \alpha$ fusion reactions but the use of the
observed O abundances through the associated contribution from p + O
spallation  appears necessary to match the observed $^6$Li abundances
of the halo stars. Fields \& Olive  adjust their model to reproduce
the solar $^6$Li which also
accounts well for the Be and B abundances of the sun, disk and halo stars.

Vangioni-Flam et al. (1999) incorporate a different mix of the
light element producing processes into their chemical evolution
model. In particular, they invoke low energy nuclei that they associate
with the acceleration of supernovae ejecta in the superbubbles created
collectively by winds from the massive stars in OB associations. (`Low energy'
refers to energies close to the threshold energies of the spallation and fusion
reactions.)  A key point about this component is that the He, C, and O
abundances of the  ejecta are considered to be much higher than in the
halo interstellar medium and, then, the dominant spallation process
 is between (say) O in the ejecta and protons in the interstellar gas
whereas in the standard picture (Fields \& Olive \cite{fields99}) the leading
process is between protons in the cosmic rays and  (say) O in the
interstellar gas.
In Fig.~\ref{fig.9}, we show predictions (dashed-dotted line)
from a model adjusted to fit the
measured $^6$Li abundances of the two halo stars. This model predicts
a $^6$Li abundance at [Fe/H] = 0 that exceeds slightly the solar
abundance.

The close correspondence between the two predictions is unlikely to
be a fair measure of the uncertainties in predicting the $^6$Li
abundance of 1~${\cal M}_{\sun}$, [Fe/H] $\simeq -0.6$ disk stars starting from
either the solar $^6$Li  abundance or the $^6$Li abundance of halo stars.
While the $^6$Li contribution from spallation by galactic high-energy
cosmic rays is rather well constrained by the observed Be abundance,
there are no comparable constraints on the 
contributions of the fusion reactions and of spallation by
low energy cosmic rays.

That the range of permissible predictions is wider than perhaps
suggested by the above two recent papers is suggested by
an earlier discussion by Yoshii et al. (\cite{yoshii97}). The
prediction shown in Fig.~\ref{fig.9} is from their Fig. 2 
\footnote{In converting
[O/H] to [Fe/H] we have adopted [O/H] = [Fe/H] + 0.5 for $\feh < -1.0$
and [O/H] = 0.5 [Fe/H] for $\feh > -1.0$}
for a model that considers high-energy cosmic rays with 
cosmic ray protons and alphas spallating interstellar C,N, and O nuclei
as well as $\alpha + \alpha$ reactions.  The cosmic
ray flux was assumed to increase with decreasing metallicity. Different
models are adopted for the halo and disk. This model predicts a rather
shallow decline of the $^6$Li abundance in the halo, and a steeper
increase in the disk. Almost all of the $^6$Li in the halo
is the product of the $\alpha + \alpha$ reactions.
The prediction fails by about 0.4 dex
to account for the solar $^6$Li abundance,
so that  the discrepancy
between prediction and observation for our disk stars might be
larger for a revised model that did reproduce the solar $^6$Li abundance.

Our old disk stars with detectable $^6$Li have Li abundances slightly
in excess of the Spite plateau. 
If, as standard models of pre-main sequence
and main sequence evolution predict, the depletion of $^7$Li
has been extremely slight, we may use the observed abundance and the
prediction that cosmic ray production of the Li isotopes gives
an isotopic ratio $^7$Li/$^6$Li $\simeq 1.5$ to estimate the 
contribution of $^6$Li from cosmic rays. Consider HD\,68284 with a
 Li abundance log$\epsilon$(Li) = 2.35. This is higher than the
Spite plateau of log$\epsilon$(Li) = 2.21 (Smith et al. 1998). On the
assumption that plateau stars have not depleted Li, the increase of Li
in HD\,68284 corresponds to Li $\simeq 60$ on the scale H =10$^{12}$.
If cosmic rays were entirely
responsible for this increase, a division $^7$Li $\simeq$ 36 and $^6$Li
$\simeq$ 24 is appropriate for a  production ratio  $^7$Li/$^6$Li $\simeq
1.5$. The Be abundance implies $^6$Li $\simeq$ 30 so that at this
metallicity spallation rather than fusion reactions may be dominant.
 The observed $^6$Li abundance is $^6$Li $\simeq$ 10. (HD\,130551
provides similar figures.) Given that $^6$Li has assuredly been
depleted to at least a modest extent, this elementary dissection of
the observed Li abundance reveals no obvious difficulty with a cosmic
ray contribution to the Li isotopes.

Recently, Ryan et al. (\cite{ryan99}) have argued  on the basis of lithium
abundances of Spite plateau stars in the range $-3.5 <$ [Fe/H] $< -2.3$
that the plateau has a metallicity dependence due to the manufacture
of the Li isotopes by cosmic rays. They consider the primordial
abundance to be log$\epsilon$(Li) $\simeq$ 2.00 at [Fe/H] = $-3.5$. 
If Smith et al.'s \teff -scale is adopted, this
abundance is raised to be about 2.12 according to 4 stars  common to both
analyses. This and Ryan et al.'s metallicity dependence predict HD\,68284
to have a Li abundance of 2.44 which is similar to the observed value of
2.35. Relative to a  plateau of 2.12, the observed abundance implies
cosmic rays have added Li in the proportion $^7$Li = 54 and $^6$Li = 36
which are consistent with our observations provided that $^6$Li has
been depleted by about 0.6 dex. At some point in the
evolution of the Galaxy, sources (presumably stellar) contributed
$^7$Li with little or no $^6$Li in order to raise the $^7$Li/$^6$Li
ratio to the solar ratio of 12.5. Inclusion of such a contribution
in the above argument reduces the $^6$Li inferred from the increase
in Li abundance over the plateau's value.

The preceding argument may be inverted: the predicted growth of $^6$Li
with [Fe/H] may be used to infer the \Liseven\ abundance. For example, 
the models proposed by Fields \& Olive, and Vangioni-Flam et al.
predict a $^6$Li abundance in the ISM at the birth of
\object{HD\,68284} and \object{HD\,130551}
of about 120, a factor of 12 greater than observed.
Assuming again a production ratio of $\sixseven = 1.5$ the attendant $^7$Li 
abundance is 180. Added to the primordial \Liseven\ abundance of 160
this implies a total \Liseven\ abundance of 340
(log$\epsilon(\Liseven) = 2.53$), a value considerably
greater than the observed value of about 200. The obvious implications are that
either the predicted $^6$Li abundance is greatly overestimated or
$^7$Li has been depleted by about 0.2 dex. In sharp
contrast, Yoshii et al.'s  prediction  is a $^6$Li
abundance of about 30 providing a total \Liseven\ abundance of 205 or
log$\epsilon$(Li) = 2.31, in excellent agreement with the observed
\Liseven\ abundances of \object{HD\,68284} and \object{HD\,130551}.
This model fails, however, to account for the meteoritic $^6$Li abundance
by a large amount.
  
\section{Concluding Remarks}

It is presently impossible to refine our conclusions because understanding
of Li-depletion, especially of $^6$Li-depletion, is poor.  
This is unfortunate as there is a tantalizing hint from the few
 detections of $^6$Li with the support of the greater suite of
upper limits to the $^6$Li abundance that the growth of the $^6$Li
abundance in the Galaxy up to the metallicity of the old disk ([Fe/H] $\sim
-0.5$) may have been very slight, a result in conflict with the most recent
models of $^6$Li chemical evolution but in agreement with a model
proposed by Yoshii et al. (\cite{yoshii97}).
This tentative suggestion is dependent on
the absence of severe $^6$Li-depletion in our disk stars. 

Further progress will need additional observations of $^6$Li in halo
and disk stars. Such observations for extremely metal-poor stars,
say [Fe/H] $< -3$, will require a high-resolution spectrograph on
an  extremely large telescope. More common instrumentation will
be adequate for expanding the sample of more metal-rich halo and
disk stars with a known $^6$Li abundance and,
 in this case, a very large sample of stars
may be needed to map out the upper envelope to the $^6$Li vs [Fe/H]
relation which with the corresponding $^7$Li vs [Fe/H] envelope
may suffice to determine empirically the evolution of the
$^7$Li/$^6$Li ratio. A by-product of the survey will
be measures of the range of $^6$Li-depletions experienced by
low mass stars of differing metallicity that will serve as
grist for the theoreticians' mills.

\begin{acknowledgements}
This research has been supported in part by the Danish Natural Science
Research Council, the US National Science Foundation
(grant 96-18414) and the Robert A. Welch Foundation of Houston, Texas.
\end{acknowledgements}

\end{document}